\documentclass[twocolumn,floatfix,aps,prd,showpacs,footinbib,amsmath,amssymb,amsfonts,superscriptaddress]{revtex4}
\usepackage{bm}
\usepackage{graphicx}
\usepackage{color}
\usepackage{dsfont}
\usepackage[normalem]{ulem}
\usepackage{hyperref}
\usepackage{mathrsfs}
\usepackage{varioref}
\usepackage[active]{srcltx}
\expandafter\ifx\csname package@font\endcsname\relax\else
\expandafter\expandafter
\expandafter\usepackage
\expandafter\expandafter
\expandafter{\csname package@font\endcsname}%
\fi
\hypersetup{
colorlinks=true,       
linkcolor=red,          
citecolor=blue,        
filecolor=blue,      
urlcolor=blue           
}

\labelformat{figure}{Fig.~#1} 
\labelformat{subfigure}{Fig.~\thefigure#1} 
\labelformat{table}{Tab.~#1} 

\def\d{{\mathrm{d}}}

\def\tr{{\mathrm{tr}}}

\def\ln{{\mathrm{ln}}}

\begin{document}

\title{Hints of quantum gravity from the horizon fluid}

\author{Bethan Cropp}
\email{bcropp, swastik, shanki@iisertvm.ac.in}
\author{Swastik Bhattacharya}
\author{S. Shankaranarayanan}
\affiliation{School of Physics, Indian Institute of Science Education and Research Thiruvananthapuram (IISER-TVM), Trivandrum 695016, India.}

\begin{abstract}
\noindent 
For many years researchers have tried to glean hints about quantum
gravity from black hole thermodynamics. However, black hole 
thermodynamics suffers from the \emph{problem of Universality} --- at 
leading order, several approaches with different microscopic 
degrees of freedom lead to Bekenstein-Hawking entropy. We attempt to 
bypass this issue by using a minimal statistical mechanical model for the 
horizon fluid based on Damour-Navier-Stokes (DNS) equation. For stationary asymptotically flat black hole spacetimes
 in General Relativity, we show explicitly that at 
equilibrium the entropy of the horizon fluid is the Bekenstein-Hawking entropy. 
Further we show that, for the bulk viscosity of the fluctuations of the horizon fluid 
to be identical to Damour, a confinement scale exists for these fluctuations, implying \emph{quantization} of the horizon area. 
The implications and possible mechanisms from the fluid point of view are discussed.

\end{abstract}

\maketitle

\noindent
Key questions in any theory of quantum gravity are {\sl what operators
  to quantize?} and {\sl what is their spectrum?}. Such a problem is
prohibitively complicated for a full theory of quantum
gravity. However, it is hoped that one can make progress when we
restrict the spacetimes under consideration. Black holes are the most
promising contenders to make such progress, as they are the some of
the simplest spacetimes.

One promising inroad into this problem has been through black hole
thermodynamics~\cite{1973-Bardeen.etal-CMP}, which has shown that the
horizon area is associated to the entropy~\cite{1972-Bekenstein-NCL,
  1973-Bekenstein-PRD, 1975-Hawking-CMP}, and hinted that this area
could be the appropriate variable to
quantize~\cite{1974-Bekenstein-NCL, 1997-Bekenstein-Proc,
  1998-Bekenstein-Arx}.  However, black hole thermodynamics now has
the {\sl problem of Universality}~\cite{2008-Carlip-Lec}; at the
leading order, several approaches using completely different
microscopic degrees of freedom lead to Bekenstein-Hawking
entropy\cite{2001-Wald-LRR}. Currently, it is not possible to identify
which are the true degrees of freedom that are responsible for the
black hole entropy~\cite{2003-Jacobson.Parentani-FP}.

Alternatively, Fluid/gravity correspondence --- projecting the
Einstein equations onto the black hole horizon leads to an equation similar in style to Navier-Stokes
\cite{1982-Damour-Proc, 1986-Price.Thorne-PRD,
  1986-Thorne.etal-Membrane,  Parikh:1997ma, 2002-Policastro.etal-JHEP,
  2008-Bhattacharyya.etal-JHEP, 2011-Padmanabhan-PRD,
  2012-Kolekar.Padmanabhan-PRD, 2012-Bredberg.etal-JHEP, Fischler:2015cma} --- can
provide a way to understand these black hole micro-states from the
microscopic degrees of freedom of the horizon fluid.  This is more
interesting as we have a better understanding about the microscopic
degrees of freedom of most fluid systems than gravity. More
specifically, given a horizon fluid equation of state, it is possible
to constrain the microscopic degrees of freedom and, hence, the {\sl
  problem of Universality} encountered in black hole thermodynamics
can be curtailed~\cite{2003-Chapline.etal-IJMPA,
  2012-Dvali.Gomez-PLB,  2014-Dvali.Gomez-EPJC, 2016-Skakala.Shankaranarayanan-IJMPD,
  2015-Lopez.etal-Arx}.  
 
Importantly, it was shown recently that
modeling of the horizon fluid as a Bose-Einstein condensate (BEC) one
may recover Bekenstein-Hawking entropy for the Schwarzschild, AdS,
 and Boulware-Deser black holes~\cite{2014-Bhattacharya.Shankaranarayanan-Arx, 2015-Lopez.etal-Arx}. 
Using the theory of fluctuations~\cite{1966-Kubo-RPP}, two of the current authors constructed~\cite{2016-Bhattacharya.Shankaranarayanan-PRD}
a statistical-mechanical description of the (number) fluctuations of  the horizon fluid and explicitly 
showed that the coefficient of bulk viscosity for the horizon fluid corresponding to 
4-D Schwarzschild black hole matches exactly with the value found from the equations
of motion for the horizon fluid~\cite{1982-Damour-Proc}. It is important to note that while the entropy 
calculation~\cite{2014-Bhattacharya.Shankaranarayanan-Arx} requires a model of the horizon fluid (i. e. BEC), 
the fluctuation dissipation analysis of the horizon fluid leading to the negative bulk-viscosity~\cite{2016-Bhattacharya.Shankaranarayanan-PRD}
is independent of the details of the horizon fluid.

The Schwarzschild black hole is the simple obvious first
test case~\cite{2014-Bhattacharya.Shankaranarayanan-Arx,2016-Bhattacharya.Shankaranarayanan-PRD}.
Since the Schwarzschild solution depends on the one parameter (the
black hole mass), it does not answer some of the important questions
like: When and why does the statistical mechanical picture of the
horizon fluid work? Does the statistical mechanical picture of the
horizon fluid and its fluctuations provide any condition on the microscopic structures of
the fluid-horizon? The purpose of this work is to answer these questions by 
extending the analyses of \cite{2014-Bhattacharya.Shankaranarayanan-Arx,2016-Bhattacharya.Shankaranarayanan-PRD} 
to D-dimensional Schwarzschild and the Kerr-Newman black holes. 

Using an appropriate quasi-local energy for the black hole \cite{1988-Kulkarni.etal-CQG,1990-Chellathurai.Dadhich-CQG,1992-Dray.etal-GRG}, and constraint equations between the thermodynamic variables leads to the correct value 
of entropy of the fluid in the above space-times. We explicitly evaluate the number density fluctuations of 
the horizon fluid in the co-rotating frame in these space-times and show that in order for the bulk
viscosity of the horizon fluid to match Damour's result~\cite{1982-Damour-Proc} 
there must be a {\sl confinement scale} i. e. fluctuations exist as bound states, as in
diverse condensed matter models \cite{1996-Delfino-NPB,1984-Witten-CMP, 1996-Shelton-PRB}. 
One of the important implications of the confinement scale is that the scale implies the quantization of 
black hole area as conjectured by Bekenstein~\cite{1974-Bekenstein-NCL, 1997-Bekenstein-Proc,
  1998-Bekenstein-Arx}.  To our knowledge, this is the first evidence that  the fluid-gravity correspondence 
provides concrete evidence for the quantization of horizon area. \\[4pt]
\noindent {\sl Horizon fluid constraint equations:} Generalizing Damour's 
calculation \cite{1982-Damour-Proc} to D-dimensional spacetime (see appendix \ref{appDNS}), 
and projecting onto the $(D - 2)$-dimensional horizon~\cite{Padmanabhan:2013nxa}, one recovers an equation similar to the Navier-Stokes 
equation describing a viscous fluid. This construction can be carried out for arbitrary null surfaces but the 
correspondence to the normal Navier-Stokes eqution is exact for the stationary spacetimes considered here. 
The fluid has pressure $(P)$, 
\begin{equation}
P= \frac{\kappa}{8\pi}=\frac{k_BT}{4},
\label{PT}
\end{equation}
where $T$ is the Hawking temperature of the black hole. In this work, we set 
$c=G=\hbar=1$ and therefore must keep $k_B$ explicit, and assume that $T>0$ corresponding to non-extremal black holes.  The horizon fluid is special in that besides the 
above constraint \eqref{PT},  it further obeys an extra constraint relating the fluid energy and temperature~\cite{2016-Skakala.Shankaranarayanan-IJMPD}.

The fluid has a naturally defined energy density, which we can integrate over the horizon area to find the total energy. This procedure bypasses the issue of choosing amongst notions of 
quasi-local energy
\cite{1978-Beig-PLA,2004-Szabados-LRR,1984-Wald-GeneralRelativity},  and naturally leads to a constraint equation relating $E$, $T$, and $A$. This energy 
is identical to one commonly used notion of quasi-local mass in the literature \cite{1988-Kulkarni.etal-CQG,1990-Chellathurai.Dadhich-CQG,1992-Dray.etal-GRG}.
See appendix \ref{app:constraint}.

For the D-dimensional Schwarzschild \cite{1963-Tangherlini-NC}, we have,	
\begin{equation}
E=M=\left(\frac{D-2}{D-3}\right)\frac{A k_BT}{4}.
\label{eq:DSchw-constraint}
\end{equation}
While the Kerr-Newman black hole, the quasi-local mass evaluated on the outer horizon 
obeys~\cite{1990-Chellathurai.Dadhich-CQG}
\begin{equation}
E=\sqrt{M^2-Q^2-a^2}=\frac{1}{2} Ak_BT. 
\label{eq:KNconstraint}
\end{equation}
Note that for $D = 4$, Eq.~\eqref{eq:DSchw-constraint} leads to Eq.~\eqref{eq:KNconstraint}.

Beside the constraints, the bulk viscosity of the horizon fluid is \emph{negative} \cite{1982-Damour-Proc}.
The coefficient of bulk ($\zeta$) and shear viscosity ($\eta$), respectively, are given by %
\begin{equation}
  \zeta=-\left(\frac{D-3}{D-2}\right)\frac{1}{8\pi}; \qquad 
	\eta=\frac{1}{16\pi}.
 \end{equation}
As mentioned earlier, two of the current authors \cite{2016-Bhattacharya.Shankaranarayanan-PRD}  
 have systematically obtained negative bulk viscosity from the fluctuations of the horizon fluid. 
 More importantly, it was shown that negative sign is due to the teleological nature of the horizon 
 \cite{1986-Price.Thorne-PRD,1986-Thorne.etal-Membrane}, as it leads to an anti-casual response to 
 infalling matter. Indeed, using a local trapping horizon instead of teleological notion of horizon one
 recovers a positive bulk viscosity \cite{2005-Gourgoulhon-PRD,2006-Gourgoulhon-PRD}. 

The above discussion clearly points that the horizon fluid corresponding to any stationary black hole is an odd system; 
the macroscopic parameters  $P, T, E$ and $A$ (volume) are not independent.  More importantly, the constraint 
equations  \eqref{PT}, \eqref{eq:DSchw-constraint}, \eqref{eq:KNconstraint} 
imply that the physical mechanism that drive the horizon fluid from an initial configuration, say, 
$P_1, T_1, E_1, A_1$ to final configuration $P_2, T_2, E_2, A_2$ can not be arbitrary and, hence, the 
fluctuations of these macroscopic quantities are also constrained.  As we will show, 
 the constraint equations play a crucial role in the derivation of the entropy and the bulk viscosity. \\[4pt]
\noindent {\sl Set up:} We list below the key ingredients in evaluating the physical quantities from the horizon fluid:
\begin{enumerate}
 \item The macroscopic properties of the horizon fluid satisfy the black hole constraints \eqref{PT} and \eqref{eq:DSchw-constraint} 
 [or \eqref{eq:KNconstraint}]. These macroscopic properties are specified by the $N$ microscopic degrees of freedom. 
 \item Following the results of \cite{2016-Skakala.Shankaranarayanan-IJMPD,1997-Kleinert-PLA}, we assume that 
there exists some temperature $T_c$ at which all $N$ microscopic degrees of freedom are 
in the ground state, and that the system remains close to the critical point $T_c$ \cite{2003-Chapline.etal-IJMPA,
  2014-Dvali.Gomez-EPJC,2014-Bhattacharya.Shankaranarayanan-Arx}.
  \item The total energy of the fluid of $N$ particles is \cite{2016-Skakala.Shankaranarayanan-IJMPD}
  \begin{equation}
E \propto N\varepsilon \propto N/r_H =N\alpha k_BT
\label{Eq:henergy}
\end{equation}
 where $\alpha$ is constant.
 \item The above energy should satisfy the constraint between energy $E$, $A$ and $T$. Following 
 \eqref{eq:DSchw-constraint} and \eqref{eq:KNconstraint}, we have
 \begin{equation}
E=Ak_BT/\gamma, 
\label{EVT}
\end{equation}
$\gamma$ is dimension dependent constant. From \eqref{Eq:henergy}, we get,
\begin{equation}
N=E/(\alpha k_BT) =A/(\gamma\alpha).
\label{NArea}
\end{equation}
\end{enumerate}

Before we proceed with the main calculations, we would like to stress the following points: 
First, our modeling of the horizon fluid is minimal. At equilibrium, $N$ contains all the information about 
the horizon fluid and all the physical variables are related to $N$. As we will show, the entropy of the 
horizon fluid is proportional to $N$ which leads to the correct Bekenstein-Hawking entropy. 
Second, to determine the transport coefficients of the horizon fluid we need to go beyond the equilibrium 
description i. e. study the fluctuations of the horizon fluid ($\delta N$)~\cite{2016-Bhattacharya.Shankaranarayanan-PRD}.
As we will show, using the theory of fluctuations~\cite{1966-Kubo-RPP}, it is possible to relate the 
bulk viscosity to the auto-correlation function of $\delta N$.\\[4pt]
\noindent {\sl Horizon fluid entropy:}  An immediate consequence of the above  set up is
the analytical prediction of the entropy of the horizon fluid using Mean field theory 
\cite{1980-Landau.Lifshitz-CourseofTheoretical}. Defining the order parameter $(\eta)$ 
of the homogeneous fluid as
\begin{equation}
\eta = \sqrt{KN} \qquad \mbox{K a positive constant} \, ,
\end{equation}
the Thermodynamic potential is given by \cite{1980-Landau.Lifshitz-CourseofTheoretical},
\begin{equation}
\Phi = \Phi_0 + a(P)\left(T- T_c \right)\eta^2+ B(P) \, \eta^4 \, ,
\end{equation}
where $a(P)$ and $B(P)$ are to be determined. Splitting into Temperature dependent and independent parts and matching the potential
$\Phi=-PA$ \footnote{One may wonder why one cannot use (\ref{PT}) and simply take the derivative w. r. t $T$ from this. 
However, as mentioned earlier,  horizon fluid is highly constrained; $A$ and $T$ are related by (\ref{eq:KNconstraint}). The mean field 
construction using the order parameter uses these constraints to obtain $a(P)$ and $B(P)$  . Furthermore, in \cite{2014-Bhattacharya.Shankaranarayanan-Arx, 2015-Lopez.etal-Arx} it
is necessary to use to the full mean field approach.} as in \cite{1980-Landau.Lifshitz-CourseofTheoretical},
we relate $P$ to $T$ via \eqref{PT} and equate the terms of order
$TA$, to get:
\begin{equation}
a=-\frac{\gamma \alpha k_B }{4K}.
\end{equation}
Unusually compared to most phase transitions, the negative value of
$a$ (as $K$, $\alpha$, and $\gamma$ must all be positive) means that
the system is in the ordered phase for $T> T_c$ an the disordered
phase for $T<T_c$ \cite{1980-Landau.Lifshitz-CourseofTheoretical}.
In the ordered phase, $\Phi$ is a minimum for 
\begin{equation}
\eta^2=KN=\frac{a(T-T_c)}{2B}.
\end{equation}

Now, if we write the entropy in the disordered and ordered phase as
$S_0$ and $S_0 + \Delta S$ respectively, one may argue that $S_0$ is
generically small as in \cite{2014-Bhattacharya.Shankaranarayanan-Arx}
(see appendix \ref{smallS}) and
\begin{equation}
\Delta S \equiv -\frac{\partial \Phi}{\partial T}= -aKN = k_B \, \frac{A}{4}.
\end{equation}
This is the first key result of this work and validates the modeling 
of the horizon fluid as a critical system. Further, we like to stress 
the following points regarding this result: (i) This is a generic result
for any $D-$dimensional stationary (spherical or axi-symmetric) black hole 
in General relativity. (ii) One of the key results from black hole thermodynamics 
is that black holes in General Relativity have an entropy $S= k_B \, A/4$. This is a key
test for modeling the horizon fluid, which has proven difficult for
several prior models \cite{2012-Dvali.Gomez-PLB, 2014-Dvali.Gomez-EPJC,
  2003-Chapline.etal-IJMPA}; the fact that this holds true for a large
class of black holes is encouraging for the success of our model of horizon fluid. \\[4pt]
 %
\noindent {\sl Bulk viscosity of horizon fluid:} {For the horizon fluid, the equilibrium state corresponds to the minimum of the Thermodynamic Potential 
$(\Phi)$. Transport phenomena in a fluid can be viewed as 
non-equilibrium processes occurring within the fluid, where the deviation from equilibrium is small. The advantage of this approach is that it is minimalistic -- indeed we may henceforth discard the assumptions about the existence of a 
phase transition (See appendix \ref{app:modelin})}
 Using the theory of fluctuations  \cite{1966-Kubo-RPP}, 
 transport coefficients can be related to autocorrelation function of number density fluctuations $(\delta N)$. Following Kubo
\cite{1966-Kubo-RPP}, the coefficient of bulk viscosity is given by:
\begin{equation}
 \zeta= \left(\frac{1}{n}\right) \frac{1}{A k_BT}\int_{-\infty}^\infty dt 
\sum\limits_a\sum\limits_b\langle J^{aa}(0)J^{bb}(t)\rangle \, ,
\label{BVformula}
\end{equation}
where, $n= Tr(\delta_{ab})$, $a, b$ run from $1, \cdots (D - 2)$ and the current $J^{ab}$ is
\begin{equation}
 J^{ab}=  \delta_{ab} \,  \delta (PV)=V \delta P \, \delta_{ab} \, .
 \label{JEq}
\end{equation}
We can write this as an entropic force --- moving the system back to equilibrium 
--- which, for the horizon fluid, takes the form:
\begin{equation}
F_{\mathrm{Th}}= P A \,  \delta A\theta(-t)
\label{eq:FTh}
\end{equation}
where $\theta(t)$ is the theta function, and enforces the anti-casual,
teleological nature of the horizon (for a detailed discussion, see
\cite{2016-Bhattacharya.Shankaranarayanan-PRD}).  The bulk viscosity
can therefore be rewritten as
\begin{eqnarray}
 \zeta &=& \frac{1}{Ak_BT}\int_{-\infty}^\infty dt \left\langle
 F_{Th}(t)F_{Th}(0)\right\rangle\nonumber \\ 
 %
%
&=& \frac{(\alpha\gamma)^2 k_B T} {16 A}\int_{-\infty}^\infty dt \left\langle \delta N(t)
 \delta N(0)\right\rangle\theta(-t) ,
\end{eqnarray}
 using \eqref{eq:FTh}, \eqref{PT} and
\eqref{Eq:henergy}. Since our interest lies in the long wavelength (fluid) limit,
 we may evaluate the viscosity from the linear response of 
the horizon fluid \cite{1966-Kubo-RPP}:
{\small 
\begin{equation}
\zeta =\lim_{\epsilon \to 0} \Im\left[\frac{(\alpha\gamma)^2 k_B T\left\langle \delta
    N^2(0)\right\rangle}{16 A}\int_{-\infty}^\infty \d t
  \exp[i(\omega-i\epsilon)]\theta(-t)\right]\\
\end{equation}
}
leading to
\begin{equation}
\zeta =-\frac{(\alpha\gamma)^2 k_B T} {16 A}\frac{\left\langle \delta
  N^2(0)\right\rangle}{\omega}.
\end{equation}
where $\omega$ corresponds to the lowest energy mode of the fluctuations that the horizon fluid support and sustain. 
Assuming that the fluctuations satisfy Maxwell-Boltzmann statistics, we get 
\begin{equation}
\left\langle \delta N^2(0)\right\rangle = \frac{4 \, A}{(\gamma \, \alpha)^2},
\end{equation}
and the bulk viscosity simplifies to
\begin{equation}
\zeta= - \frac{k_B \, T}{4\, \omega}.
\label{eq:zeta1}
\end{equation}
As this is the second main result of this work we would like to stress the following points: \\
\noindent (i) Bulk viscosity is independent of the constants $\alpha, \gamma$; these can only be 
determined with the knowledge of the microscopic theory.  This is consistent as the fluid description 
does not require complete knowledge of the microscopic degrees of freedom. \\
\noindent (ii) Bulk viscosity depends on the horizon fluid temperature and the 
lowest energy mode of the fluctuations. To go about determining the lowest energy mode, it 
may be important to get a physical insight.  Let us consider fluctuations that cause a 
change in the horizon fluid  area from $A$ to $A + dA$. For the horizon fluid corresponding to Schwarzschild black hole, the 
change in area is only due to the change in the quasi-local energy.  In Ref. \cite{2016-Bhattacharya.Shankaranarayanan-PRD} 
this was done by picking the largest wavelength to be the circumference of the black hole. However, for generic
black holes like Kerr-Newman, the change in the area can also be due to the change in the electrostatic energy 
or quasi-local energy $\kappa \times dA$ \cite{1984-Wald-GeneralRelativity}. Using the fact that the lowest 
energy modes are adiabatic (slowly evolving), and that the energy and area 
are strongly constrained,  
the minimum energy change can be related to a minimum change in area through  \eqref{Eq:henergy} and \eqref{NArea}:
\begin{equation}
\Delta E = (\Delta N) \alpha T \propto (\Delta A)T \, ,
\label{EArel}
\end{equation}
and, thus, using \eqref{eq:KNconstraint} the minimum energy mode for any stationary black hole is 
\begin{equation}
\omega = \Delta E = \left(\frac{D-2}{D-3}\right)\, \Delta A \, \frac{T}{4} \, .
\label{eq:omegamin1}
\end{equation}
This brings us to the key result of this work and indicates 
that the existence of the lowest energy mode of fluctuation 
($\delta N$) of the horizon fluid relates to the \emph{minimum 
change} or \emph{quantization} of the horizon area. We will now 
make several comments about the result and embed them in a 
broader context: 

First, the area quantization has long been a key result of both horizon
thermodynamics and various quantum gravity proposals. The first
attempt goes back to Bekenstein \cite{1973-Bekenstein-PRD}, who uses
the fact that particles entering the black hole cannot have zero size,
to find the minimum area increase when one is absorbed by a
Kerr-Newman black hole. This result is very general
\cite{2016-Skakala.Shankaranarayanan-IJMPD}, and in D-dimensions leads
to
\begin{equation}
\Delta A_\mathrm{min} =8\pi \ell_P^{D-2},
\label{Amin}
\end{equation}
where we have briefly re-instituted an explicit Planck length
($\ell_P$) for clarity.  Substituting Bekenstein's minimum area in \eqref{eq:omegamin1},
we get,
\begin{equation}
\omega=  2\pi\left(\frac{D-2}{D-3}\right)  \, k_B \, T \, .
\end{equation}
Substituting the above form of minimum energy mode in \eqref{eq:zeta1}, we get
\begin{equation}
\zeta = -\left(\frac{D-2}{D-3}\right)\frac{1}{8\pi} \, .
\label{eq:zeta2}
\end{equation}
It is important to note that this is the expression for bulk viscosity for all asymptotically 
flat space-times in all dimensions and matches exactly the expression from the DNS equation 
(see appendix \ref{appDNS}). This is a \emph{non-trivial} result, using fluctuation-dissipation \cite{1966-Kubo-RPP} 
and the fact that the horizon is anti-causal \cite{2016-Bhattacharya.Shankaranarayanan-PRD}, it is not possible to obtain the 
known bulk viscosity without invoking area quantization.

Second, so far, it has not been necessary to place a value on $\alpha$. However, if we take 
$\Delta N_\mathrm{min}=1$ using \eqref{Amin} , 
\begin{equation}
\alpha = 8 \pi \, ,
\label{eq:alpha}
\end{equation}
corresponding to \cite{1973-Bekenstein-PRD}. Naturally, there have been several alternative 
proposals from counting black hole microstates or Bohr's principle with quasi-normal modes
\cite{1995-Bekenstein.Mukhanov-PLB, 1998-Hod-PRL}, and arguments from
Loop quantum Gravity \cite{2003-Dreyer-PRL}, for various values of
$\alpha$. Note that all these proposals give answers of the same order
of magnitude (see \cite{2014-Skakala.Shankaranarayanan-PRDa, Kothawala:2008in}).

Third, why should the existence of the long wavelength mode imply quantization of horizon area? 
In other words, why does the IR cutoff (lowest energy mode) have anything to do with the Planck-scale 
Physics? One of the main assumptions in our work is that the horizon fluid is described by $N$ 
microscopic degrees of freedom. While we do not have any information about these microscopic 
structures, existence of the largest length scale of the fluctuation $\lambda$ leads to the fact that 
$\lambda^{D-2} \simeq N \ell_P^{D-2}$ where $\ell_P$ physically refers to the separation of the microscopic 
structures or the size of the structures themselves. In the language of \emph{Wheeler's It from Bit} 
\cite{1992-Wheeler-Proc}, Bits are the $N$ microscopic structures of the horizon fluid. 

Fourth, why should the minimum area condition give a minimum $\omega$ from the
horizon fluid? Consider \eqref{Eq:henergy}: with $\alpha$ as given in \eqref{eq:alpha}. 
Adding one extra particle to the ground state of the condensate
is the minimum energy we can add to the system, and corresponds
exactly to adding a minimum of extra area. As this energy is larger
than or equal to the energy of the mode spanning the circumference of
the horizon, the adding of one extra horizon fluid particle is the
minimum energy. This physically motivates picking the proportionality factor as 
unity between $\omega$ and $\Delta E$ in Eq. (\ref{eq:omegamin1}).

Fifth, what does the minimum energy mode physically correspond in the horizon fluid? 
The existence of a minimum energy mode implies a confinement scale i.e. density fluctuations exist 
as bound states \cite{1996-Delfino-NPB,1984-Witten-CMP, 1996-Shelton-PRB}.  In other words, 
this implies that to add one microscopic constituent to the horizon fluid will cost that much energy.
In the language of statistical mechanics, the minimal energy mode can be treated as a chemical 
potential corresponding to the excitations in the fluid. In appendix \ref{chempot}, 
we have constructed an explicit model that shows $\mu \propto T$. 

Sixth, the status of the third law is unclear in black hole thermodynamics: while the process version of the 
third law seems to hold,
the Planck version, stating that the entropy goes to zero as the
temperature does, clearly does not hold for black holes \cite{2001-Wald-LRR}.  It
naturally follows from the existence of a positive chemical potential (see  appendix \ref{chempot}) that the system
retains a residual entropy, analogous to that of other condensed
matter systems, and probably implying a degeneracy in the ground state of the system. The large entropy of 
black holes at
 low temperature can possibly be viewed as residual entropy.  

Finally, the fact the horizon fluid has a confinement scale signals the occurrence 
of a mass gap.  There has been 
a long history of possible connection between the General relativity 
and Yang-Mills theory, in which mass gap is known to occur. Specifically, it has been shown that 
four-graviton amplitudes with one loop in Einstein gravity 
are similar to products of integrands appearing in gauge theory~\cite{1987-Kawai.Lewellen.Tye-NPB,2002-Bern-LRR,2007-Ananth.Theisen-PLB}. 
We suggest here that it might be possible to understand the
connection to Yang-Mills through the presence of a mass gap for the horizon fluid.

There are a number of ways in which one could extend and generalize this result:
These calculations were carried out using the natural split between space and time given to us by the fluid picture, 
but one can make alternative choices, and it may be elucidating to check what changes in the 
analysis when performed in a different frame. Possible extensions to other theories of gravity, including supersymmetric 
theories where we have a better understanding of microscopic degrees of freedom could be elucidating.  
Further exploration into the nature of the fluid and
the mechanism generating the IR cut-off may provide further avenues
for such insights. The statistical mechanical picture of
the horizon fluid developed over the past few years can provide a new
and exciting window into the nature of gravity, and hints towards
quantum gravity.

Coming back to the problem of Universality, it is clear that reproducing the black hole entropy is only one test of a microscopic 
model of black hole physics, and that the viscosities in the  DNS equation can help us choose between the multitude of different scenarios.
\\[4pt]
{\sl Acknowledgements:} The authors wish to thank Kinjalk Lochan for discussions, and  Stefano Liberati and 
T. Padmanabhan for discussions and comments on earlier draft of the manuscript. The work is supported under DST-Max Planck Partner 
Group on Cosmology and Gravity. 

\appendix

\section{Damour-Navier-Stokes equation in arbitrary dimensions}
\label{appDNS}

Closely following the procedure adopted in \cite{2011-Padmanabhan-PRD}, take a
$D-$dimensional spacetime. We will use small Latin letters for indices
running over all $D$ dimensions, Greek letters for $(D - 1)-$dimensional null
surface, and capital Latin for $(D-2)$ space-like dimensions.

Start with the Einstein equations,
\begin{equation}
R_{ab} -\frac{1}{2}g_{ab} = 8\pi T_{ab}, 
\end{equation}
and consider a null surface, which is therefore traced out by null
geodesics. These are described by the geodesic equation (in this case
with a non-affine parameterization),
\begin{equation}
l^a\nabla_a l_b=\kappa l_b.
\end{equation}
We can, without loss of generality, choose coordinates such that
\begin{equation}
l= \partial_t +v^A\partial_A; \qquad  l^a = (1, v^A, 0).
\end{equation}
We can also, for convenience, construct another null vector, k, such that
$k\cdot l = -1$.

The metric on the $(D-2)-$dimensional surface, denoted by
$q_{AB}$, for which
\begin{eqnarray}
ds^2 &=& q_{AB}(dx^A -v^A dt)(dx^B -v^B dt), \nonumber \\
q_{ab} &=& g_{ab} + l_ak_a + l_b k_a.
\end{eqnarray}
It can explicitly be seen that
\begin{equation}
q_{ab}l^b = q_{ab}k^b =0.
\label{qabeqns}
\end{equation} 

The Damour-Navier-Stokes (DNS) equation is a consequence of the contracted
Codazzi equation formed with $l$ and $q_{AB}$,
\begin{eqnarray}
\label{codazzifirst}
R_{mn}l^m q^n_a &\equiv& R_{mA}l^m =(\frac{1}{2}g_{ab}+ 8\pi
T_{ab})l^mq^n_a \nonumber \\ &=& 8\pi T_{ab}l^mq^n_a.
\end{eqnarray}
Here the last equality is through (\ref{qabeqns}). 
We re-write the LHS of \ref{codazzifirst} as
\begin{equation}
R_{mA}l^m = R_{\mu A}= \nabla_\mu (\nabla_A l^\mu) -\partial_A(\nabla_\mu l^\mu).
\label{codazzi}
\end{equation}
We expand both terms of the RHS of \ref{codazzi} in terms of expansion, shear and the
velocity $v_A$.

Define 
\begin{equation}
\chi^\beta_\alpha= \nabla_\alpha l^\beta
\end{equation}
and
\begin{equation}
\omega_\alpha= \chi^0_\alpha,
\end{equation}
which is the energy-momentum vector of the horizon fluid

As in equation (21) of \cite{2011-Padmanabhan-PRD} we expand out the
second term of the RHS,
\begin{eqnarray}
\partial_A(\nabla_\mu l^\mu) &=& \nabla_A l^A + \nabla_0 l^0 \nonumber \\
&=& \theta + \omega_A v^A +\omega_0 \nonumber \\
&=& \theta + \kappa, 
\label{expansionkappa}
\end{eqnarray}
and is independent of $D$.

The other term of equation (\ref{codazzi}),
\begin{equation}
\nabla_\mu (\nabla_A l^\mu)= \nabla_\mu \chi^\mu_A, 
\end{equation}
can be evaluated by taking a frame where we neglect Christoffel
symbols (see discussion in \cite{2011-Padmanabhan-PRD}), so that
\begin{equation}
\nabla_\mu \chi^\mu_A =\partial_\mu \chi^\mu_A=\partial_0 \omega_A +\partial_B \chi^B_A.
\end{equation}
Now use the fact that
\begin{equation}
\chi_{AB}= \Theta_{AB}+\omega_A v_B, 
\end{equation}
and we can further split $\Theta_{AB}$ into trace and traceless parts
\begin{equation}
\Theta_{AB}= \sigma_{AB} +\left(\frac{1}{D-2}\right)\, \theta \delta^A_B.
\end{equation}
Note the dimensional dependent prefactor, ensuring
$\tr(\Theta)=\theta$.

Putting all this together into equation \eqref{codazzi}
\begin{eqnarray}
R_{mA}l^m &=& (\partial_0 + v^B\partial_B)\omega_A -\partial_A(\kappa
+ \theta) \nonumber \\ &+&
\partial_B\left(\sigma^B_A+\frac{1}{D-2}\theta\delta^B_A\right).
\end{eqnarray}
So that
\begin{eqnarray}
8\pi T_{mA}l^m &=&(\partial_0 + v^B\partial_B)\omega_A+ \partial_B
\sigma^B_A \nonumber \\ &-&\partial_A \kappa
-\frac{D-3}{D-2}\partial_A \theta,
\end{eqnarray}
or equivalently 
\begin{eqnarray}
-\frac{(\partial_0 + v^B\partial_B)\omega_A}{8\pi}&=& -\frac{\partial}{\partial
  x^A}(\frac{\kappa}{8\pi}) +2 \frac{1}{16\pi}\sigma^B_{A |
  B} - l^aT_{aA}\nonumber\\ 
	&-&\left(\frac{D-3}{D-2}\right)\frac{1}{8\pi}\frac{\partial\theta}{\partial
  x^A}.
\label{NavStok}
 \end{eqnarray}
where $-\omega_A/(8\pi)$ is the momentum density of the fluid. 

We thus arrive at a Navier-Stokes equation with pressure:
\begin{equation}
P=\frac{\kappa}{8\pi}=\frac{k_BT}{4},
\end{equation}
shear viscosity
\begin{equation}
\eta= \frac{1}{16\pi},
\end{equation}
and bulk viscosity
\begin{equation}
\xi=-\left(\frac{D-3}{D-2}\right)\frac{1}{8\pi}.
\end{equation}

As noted in \cite{2011-Padmanabhan-PRD}, there are some differences between the DNS and the usual Navier-Stokes (NS) equation in the case of a general null surface. Firstly, the shear vector is not constructed solely from the velocity of the fluid, having an additional term. In the case of a stationary horizon, the extra term, $\partial_0 q_{AB}$ goes to zero and the shear tensor has the usual form. Secondly, in the DNS equation there is a Lie derivative rather than a convective derivative in the normal fluid case, consisting of an additional term of the form $\omega_B D_A v^B$. Again, for stationary spacetimes, this term goes to zero as the velocity is a constant. 

In deriving (\ref{NavStok}) we have worked with a particular coordinate system. One may wonder how the equation is affected when changing to another coordinate system, in particular changing to a different time coordinate. Examining (\ref{NavStok}), one may see that the difference is that the first term $(\partial_0 + v^B\partial_B)\omega_A$ will change. As long as we work with the class of transformations such that we do not introduce an explicit time dependence in the definitions of the spacelike-coordinates, $v^A$ will remain constant, and the  $\omega_B D_A v^B$ term will not reappear. The DNS equation will transform in the same way as changing coordinates in the NS equation, in a similar manner to looking at a normal laboratory fluid in a boosted frame.

We could consider this from a different angle, and ask, for a general null surface, is our analysis of the bulk viscosity coefficient independent of the boost frame chosen? We answer here the question in the affirmative and sketch an argument below why. 
For  \emph{processes concerning only the bulk viscosity}, we need only consider the part 
of the free energy that comes from its volume. Thus the part of the free energy that is dependent on the 
velocity of the horizon-fluid does not play any role in the processes solely involving changes in the bulk of 
the fluid. The extra term that comes due to going over to a boosted inertial frame, is for a general null surface, 
$\omega_B D_A v^B$, but this term does not affect calculations relating to the bulk viscosity. The change in 
the volume of the cross section of the horizon does affect $\omega$. This 
can be seen from the the fact that the rate of the change of the volume, the scalar $\theta$ is given by the divergence 
of the 
null normal to the horizon. This means as we boost the frame, $\omega$ can change
 but $\theta$ would remain the same. Therefore the term in DNS equation, which makes it 
different from the standard Navier-Stokes equation need not be considered when looking at processes 
involving change in the bulk of the fluid only. Thus whichever frame we do the calculation in, it would 
give us the same result as when we calculated the free energy in the frame comoving with the horizon-fluid. 
So our analysis does not depend on the boost.

\section{Constraint equations and horizon fluid energy}
\label{app:constraint}

To understand thermodynamic relations for the horizon fluids, we must
first define the appropriate notion of the fluid energy. There are
several definitions of energy in general relativity~\cite{1978-Beig-PLA,2004-Szabados-LRR,1984-Wald-GeneralRelativity}. 
Here we want a quasi-local notion to associate to the black hole horizon. 
Various competing definitions exist, but our task is simplified
by the fact that we have a naturally defined fluid energy density on
horizon, given by $\frac{\omega_0}{8\pi}$ where
\begin{equation}
\omega_0 \equiv \nabla_0 l^0 = \kappa
\end{equation}
and the equality follows from \eqref{expansionkappa}, noting that as
the $v_A\omega_B$ is antisymmetric its trace is automatically zero.

The total energy of the horizon is now simply
\begin{equation}
E= \frac{1}{8\pi}\int \kappa \, \d A
\end{equation}
where both $\kappa$ and $A$ are functions of all the black hole
variables ($M, Q, a$). This gives the
natural value of $M$ in the $D-$dimensional Schwarzschild.

In fact, from Eqn's (12.5.33) - (12.5.37) in
\cite{1984-Wald-GeneralRelativity}, one can see that this is exactly

\begin{equation}
E=\frac{1}{8\pi}\int \kappa \d A = \frac{1}{8\pi}\oint_S \d S_{\mu \nu} \nabla^\mu l^\nu; \qquad l= \partial_t +v^A\partial_A
\end{equation}
when $v_A$ is the rotational velocity at the horizon, making $l$
the standard combination considered when evaluating, e.g. the surface
gravity of black holes.  This corresponds to a quasi-local mass
frequently used in the literature
\cite{1988-Kulkarni.etal-CQG,
  1990-Chellathurai.Dadhich-CQG}, evaluated on the horizon.  This is
almost equal to the well-known Komar mass, the difference being the
replacement $\partial_t \to l$.  Physically this corresponds to a quasi-local
energy in the co-rotating frame \cite{1992-Dray.etal-GRG}, the
appropriate choice for the fluid that co-rotates with the horizon, as
ours does.

Explicitly, for the Kerr-Newman we find
\begin{equation}
E=\frac{r_+-r_-}{2}.= \sqrt{M^2-Q^2-a^2}
\end{equation}

The black hole is an odd system in many ways, as $P, T, E$ and $A$ are
not independent. Instead they obey $P=T/4$ and an extra constraint
equation. The form of this equation varies depending on the class of
black holes. We derive this relation for the
$D$-dimensional Schwarzschild black hole and the (4D) Kerr-Newman
black hole.

\subsection{D-dimensional Schwarzschild}
The D-dimensional Schwarzschild, also known as a the
Schwarzschild-Tangherlini black hole \cite{1963-Tangherlini-NC}, has
the form
\begin{eqnarray}
\d s^2&=&-\left(1-\left(\frac{r_H}{r}\right)^{D-3} \right)\d t^2 \nonumber \\
&+&\frac{\d r^2}{1-\left(\frac{r_H}{r}\right)^{D-3}} +r^2 \d \Omega^2_{D-2},
\end{eqnarray}
where
\begin{equation}
\Omega_{D-2}=\frac{2\pi^{\frac{D-1}{2}}}{\Gamma\left(\frac{D-1}{2}\right)}.
\end{equation}
The horizon ``area" is now given by
\begin{equation}
\label{Darea}
A= \Omega_{D-2}r_H^{D-2}.
\end{equation}
and the standard temperature is 
\begin{equation}
T=\frac{D-3}{4\pi k_B r_H}.
\end{equation}
Here the horizon radius is related to the mass of the black hole by
\begin{equation}
\label{Dradius}
r_H=\left(\frac{16\pi M}{(D-2)\Omega^2_{D-2}}\right)^{\frac{1}{D-3}}.
\end{equation}
This $M$ is the energy of the horizon fluid, so
\eqref{Darea}--\eqref{Dradius} combine to give a constraint equation
\begin{equation}
E=\left(\frac{D-2}{D-3}\right)\frac{k_B A T}{4}.
\end{equation}

\subsection{Kerr-Newman}

The charged, rotating black hole,
\begin{eqnarray}
\d s^2 &=& -\left(\d t -a\sin^2 \theta\d
\phi\right)^2\frac{\Delta}{\rho^2}  - \left(\frac{\d r^2}{\Delta}+\d \theta^2\right)\rho^2 \nonumber \\
 &+& \left((r^2+a^2)\d \phi - a \d t \right)^2\frac{\sin^2\theta}{\rho^2} 
\end{eqnarray}
where
\begin{equation}
\rho^2 = r^2 +a^2\cos^2\theta; \quad \Delta= r^2-2Mr+Q^2+a^2.
\end{equation}
has inner and outer horizons given by
\begin{equation}
r_\pm =M+\sqrt{M^2 - Q^2 -a^2}
\end{equation}
We will only be concerned by the physically relevant outer horizon.

The horizon area and black hole temperature can be easily seen to be
\begin{equation}
A=4\pi(r_+^2+a^2); \qquad T= \frac{1}{2\pi}\frac{r_+ - r_-}{2k_B(r_+^2 +a^2)}.
\end{equation}
And the energy on horizon is \cite{1990-Chellathurai.Dadhich-CQG}
\begin{equation}
E=\frac{r_+-r_-}{2}.
\end{equation}
Combining these results we can see that
\begin{equation}
E=\frac{1}{2}k_B AT. 
\label{KNconstraint}
\end{equation}
 Note that this energy is associated here first and foremost to the horizon fluid, rather than the energy contained within the horizons, and therefore different to the similar relation in \cite{Padmanabhan:2013nxa}. 
This is identical to the expression for the 4D expression
for the D-dimensional black hole, though the forms of $A$ and $T$ are very different.
These constraint equations play a crucial role in the derivation
of the entropy and the bulk viscosity.

\section{Chemical Potential and IR cutoff for the density waves in the Horizon Fluid}
\label{chempot}

Here we shall point out that there is another way to view the physical condition
imposed by the IR cutoff in the frequency of the perturbations. Consider density wave excitations, which are bosons.
The presence of such a cutoff signals the existence of a positive chemical potential,
$\mu$ for these excitations. In fact the presence of a
positive chemical potential ensures that there can be no excitation
with frequency $\Omega$ such that $\Omega<\mu$. This can be seen by considering the
expression for the occupation number of such excitations with
frequency $\Omega$. Let us denote the occupation number by
$n_\Omega$. Then, for degeneracy $g_\Omega$,
\begin{equation}
n_{\Omega}= \frac{g_{\Omega}}{\exp{(\mu+\beta\Omega)}-1}. \label{occupnno}
\end{equation}
It is seen from \eqref{occupnno} that $n_\Omega>0$ only if
$\Omega>\mu$. In our case, $\mu=\Omega_{IR}$, hence density waves have
a frequency minimum given by that value.

There are two important physical implications of the fact $\mu\propto
T$ for the horizon fluid.

\begin{itemize}

\item The fluctuations in the area of the horizon fluid are
  quantized \cite{1980-Landau.Lifshitz-CourseofTheoretical}.
   To see how this comes about, let us note
  that from thermodynamic relations, one can write,
\begin{equation}
 \mu \delta n= -T\delta S. \label{MuT}
\end{equation}
If $\mu= CT$, then we have, $\delta S= C\delta n$. Since, $\delta S= \frac{1}{4}\delta A$, it follows that, 
\begin{equation}
 \delta A= 4C\delta n, \label{AreaQ}
\end{equation}
which tells us that area is quantized in terms of an unit, whose value
is given by $4C$. Here, $C$ is positive. From \eqref{MuT}, we see that
this makes $\delta S<0$. 

The change in entropy is negative because we are dealing with the system moving towards equilibrium, the higher entropy state, only obtaining equilibrium in the future. So if there are fluctuations in the system, i.e. it is excited, then it has less
entropy.  {\it This shows us that the area quantization can also be
  viewed as a consequence of
  $\Omega_{IR}\propto T$.}

\item The entropy density of the perturbations of the horizon fluid
  has a constant term. Like the previous result, this one is
  also a direct consequence of $\mu\propto T$. A similar case is that
  of liquid Helium $\mathrm{He}^3$ close to zero temperature (see
  chapter 21, in \cite{2012-Lifshitz-Perspectives}).  This constant
  entropy term is referred to as the residual entropy in the
  literature \cite{1999-Atkins.dePaula-physicalchemistry}. Typically
  it occurs because at very low temperatures, the systems have
  degeneracy at the configurational level  \cite{1935-Pauling-JACS,1987-Chow.Wu-PRB}. 
  Well-known systems with this property are the Carbon Monoxide gas, where the residual
  entropy is present due to a degeneracy in the molecular
  arrangement~\cite{1999-Atkins.dePaula-physicalchemistry}, and the
  liquid Helium ${\mathrm{He}}^3$, where the residual entropy results
  due to the disordered orientation of the nuclear spins of He
  atoms~\cite{2012-Lifshitz-Perspectives}. In analogy with the known
  cases, residual entropy can be thought of as due to some unknown
  configurational degeneracy in the horizon fluid. These
  configurations correspond to unknown physics presumably
  corresponding to some kind of underlying discrete structure of black
  holes.
   \end{itemize}

\section{Entropy of the disordered phase}
\label{smallS}

This discussion will follow very closely that of
\cite{2014-Bhattacharya.Shankaranarayanan-Arx}.  The horizon fluid has
a low energy in the disordered phase, as in this phase $T < T_c$. For
simplicity, we assume that the system can be described by a scalar
field, $\phi$, but this is not a key feature of the derivation.

In the disordered phase, the partition function can be expressed by
\begin{eqnarray}
Z&=& \int_{p_{_{\mathrm{Low}}}} D \phi \exp{\left[-\frac{\beta}{2}\int \d^2x \phi(x)(-\nabla^2+m^2)\phi(x)\right]} \nonumber \\
&=& \mathrm{Det}\left[\beta(-\nabla^2+m^2)\right]^{-1/2},
\end{eqnarray}
where $p_{_{\mathrm{Low}}}$ indicates we are only integrating over
low-momentum modes.  Standard calculations lead to:
\begin{equation}
\ln Z= A\int_{p_{_{\mathrm{Low}}}} \frac{\d^2p}{(2\pi)^2} \ln{\left[\beta(-\nabla^2+m^2)\right]}.
\end{equation}
If $p_{_{\mathrm{Low}}}$ is zero, this entropy would also be
zero. However, this will not be zero, due to the presence of the
minimum area change. Then
\begin{equation}
\ln Z= A \ln(m^2\beta)\frac{p_\mathrm{min}^2}{(2\pi)^2} \approx  A \ln(\beta)\frac{p_\mathrm{min}^2}{(2\pi)^2} .
\end{equation}
Now, using $p_\mathrm{min}=\omega$
\begin{equation}
\ln Z = A(k_BT)^2 \left(\frac{D-2}{D-3}\right)^2 \ln\left(\frac{1}{k_BT}\right) \, ,
\end{equation}
which is small. In the case of Kerr, for instance, when $a \ll M$, $\ln
Z \sim \ln(r_+)$ as in the Schwarzschild case, while for $a \approx
M$, $T^2$ dominates over $\ln(1/T)$ and $S_0$ goes to zero.

\section{Model-independent derivation of coefficient of bulk viscosity}
\label{app:modelin}

In the main body of the paper, we have explicitly used $N \propto A$, which is a natural consequence of the horizon fluid being in its ground state. Here we will show that the Fluctuation-Dissipation analysis of the bulk viscosity of the 
horizon-fluid is independent of such details, and of the microscopic the structure of the fluid, as had been 
noted in \cite{2016-Bhattacharya.Shankaranarayanan-PRD}.

The horizon-fluid has two constraints, between $P$ and $T$; and relating $E$, $A$ and $T$. We see explicitly that the volume of the horizon-fluid contributes to its free energy. 
For the processes corresponding only to the changes in the bulk, only this part of the free energy need be 
considered, so only this part will concern us. Thus, we can make the choice to go over to the rest frame of the fluid and write down the free energy relevant 
for the bulk viscosity. It is important to note that the analysis can be performed 
where the 'kinetic' energy of the fluid constituents also contributes to the total energy 
$E$ (as discussed in section I), as we only need the relevant part of the energy. 

For the highly constrained system of the horizon-fluid, it turns out that $(P,E,T,A)$ are all related to each 
other to the extent that the free energy 
can be expressed in terms of the variation of a single variable. This univariability of the horizon-fluid was seen 
for the 
Schwarzschild black hole in \cite{2016-Bhattacharya.Shankaranarayanan-PRD} and has now been found to persist in the generalizations made here. 
The variable $A$ is a good physical choice and the free energy of the horizon-fluid can be expressed in 
its terms. For example, this  is what we have done in the specific model for the fluid, and can be seen from the relations 
\begin{eqnarray}
 \Phi&=& \Phi_0+a(P)(T-T_c)\eta^2+B(P)\eta^4;\\  
\eta&=& \sqrt{\frac{k}{2\alpha}}\sqrt{A}. \label{phiLT}
\end{eqnarray}
Here $a$, $B$ are parameters as defined in the main body of the paper.  

To describe transport processes, we need to consider the fluid at a state slightly away 
from equilibrium, the minimum value of the free energy. We can express the fluctuation in free energy as,
\begin{equation}
\delta\Phi= \frac{1}{8} (T-T_c)\frac{\delta A^2}{A}. \label{deltaphi}
\end{equation}
Here we note that whichever variable we choose to express the free energy in, the lowest order fluctuation 
in the free energy of the horizon-fluid has to be quadratic in that variable. \emph{This quadratic dependence, $\delta \Phi \propto \delta X^2 $ is the key relation we need and does not depend on the specific relation of (\ref{phiLT}), merely coming from the fact that the free energy is proportional to the convienent varibale we are working with} . Now any fluid is composed of 
microscopic dof, whose total number, $N$ however, can be treated as a macroscopic variable. So it is 
possible to express the free energy in terms of $N$ as well. Hence the change in the free energy, 
$\delta\Phi$ is quadratic in $\delta N$, the fluctuation in the variable $N$. In that case, it follows that 
$\delta N\propto\delta A$. It is to be emphasized here that we obtain this result here without assuming 
any proportionality between $N$ and $A$ as in the case of the BEC model of the horizon-fluid. 

One can now proceed in the same way 
as shown in the paper and evaluate the coefficient of bulk viscosity as shown in the main body of the paper.


\begin{thebibliography}{52}
\expandafter\ifx\csname natexlab\endcsname\relax\def\natexlab#1{#1}\fi
\expandafter\ifx\csname bibnamefont\endcsname\relax
  \def\bibnamefont#1{#1}\fi
\expandafter\ifx\csname bibfnamefont\endcsname\relax
  \def\bibfnamefont#1{#1}\fi
\expandafter\ifx\csname citenamefont\endcsname\relax
  \def\citenamefont#1{#1}\fi
\expandafter\ifx\csname url\endcsname\relax
  \def\url#1{\texttt{#1}}\fi
\expandafter\ifx\csname urlprefix\endcsname\relax\def\urlprefix{URL }\fi
\providecommand{\bibinfo}[2]{#2}
\providecommand{\eprint}[2][]{\url{#2}}

\bibitem[{\citenamefont{Bardeen et~al.}(1973)\citenamefont{Bardeen, Carter, and
  Hawking}}]{1973-Bardeen.etal-CMP}
\bibinfo{author}{\bibfnamefont{J.~M.} \bibnamefont{Bardeen}},
  \bibinfo{author}{\bibfnamefont{B.}~\bibnamefont{Carter}}, \bibnamefont{and}
  \bibinfo{author}{\bibfnamefont{S.~W.} \bibnamefont{Hawking}},
  \bibinfo{journal}{Commun. Math. Phys.} \textbf{\bibinfo{volume}{31}},
  \bibinfo{pages}{161} (\bibinfo{year}{1973}).

\bibitem[{\citenamefont{Bekenstein}(1972)}]{1972-Bekenstein-NCL}
\bibinfo{author}{\bibfnamefont{J.~D.} \bibnamefont{Bekenstein}},
  \bibinfo{journal}{Nuovo Cimento Lett} \textbf{\bibinfo{volume}{4}},
  \bibinfo{pages}{737} (\bibinfo{year}{1972}).

\bibitem[{\citenamefont{Bekenstein}(1973)}]{1973-Bekenstein-PRD}
\bibinfo{author}{\bibfnamefont{J.~D.} \bibnamefont{Bekenstein}},
  \bibinfo{journal}{Phys. Rev.} \textbf{\bibinfo{volume}{D7}},
  \bibinfo{pages}{2333} (\bibinfo{year}{1973}).

\bibitem[{\citenamefont{Hawking}(1975)}]{1975-Hawking-CMP}
\bibinfo{author}{\bibfnamefont{S.~W.} \bibnamefont{Hawking}},
  \bibinfo{journal}{Commun. Math. Phys.} \textbf{\bibinfo{volume}{43}},
  \bibinfo{pages}{199} (\bibinfo{year}{1975}).

\bibitem[{\citenamefont{Bekenstein}(1974)}]{1974-Bekenstein-NCL}
\bibinfo{author}{\bibfnamefont{J.~D.} \bibnamefont{Bekenstein}},
  \bibinfo{journal}{Lettere al Nuovo Cimento (1971-1985)}
  \textbf{\bibinfo{volume}{11}}, \bibinfo{pages}{467} (\bibinfo{year}{1974}).

\bibitem[{\citenamefont{Bekenstein}(1997)}]{1997-Bekenstein-Proc}
\bibinfo{author}{\bibfnamefont{J.~D.} \bibnamefont{Bekenstein}}, in
  \emph{\bibinfo{booktitle}{{Proceedings of 8th Marcel Grossmann meeting, Israel}}} (\bibinfo{year}{1997}), pp.
  \bibinfo{pages}{92--111}, \eprint{gr-qc/9710076},

\bibitem[{\citenamefont{Bekenstein}(1998)}]{1998-Bekenstein-Arx}
\bibinfo{author}{\bibfnamefont{J.~D.} \bibnamefont{Bekenstein}}, in
  \emph{\bibinfo{booktitle}{{9th Brazilian School of Cosmology and Gravitation, Brazil}}}
  (\bibinfo{year}{1998}), \eprint{gr-qc/9808028},

\bibitem[{\citenamefont{Carlip}(2009)}]{2008-Carlip-Lec}
\bibinfo{author}{\bibfnamefont{S.}~\bibnamefont{Carlip}},
  \bibinfo{journal}{Lect. Notes Phys.} \textbf{\bibinfo{volume}{769}},
  \bibinfo{pages}{89} (\bibinfo{year}{2009}), \eprint{0807.4520}.

\bibitem[{\citenamefont{Wald}(2001)}]{2001-Wald-LRR}
\bibinfo{author}{\bibfnamefont{R.~M.} \bibnamefont{Wald}},
  \bibinfo{journal}{Liv. Rev. Rela.} \textbf{\bibinfo{volume}{4}},
  \bibinfo{pages}{6} (\bibinfo{year}{2001}), \eprint{gr-qc/9912119},

\bibitem[{\citenamefont{Jacobson and
  Parentani}(2003)}]{2003-Jacobson.Parentani-FP}
\bibinfo{author}{\bibfnamefont{T.}~\bibnamefont{Jacobson}} \bibnamefont{and}
  \bibinfo{author}{\bibfnamefont{R.}~\bibnamefont{Parentani}},
  \bibinfo{journal}{Found. Phys.} \textbf{\bibinfo{volume}{33}},
  \bibinfo{pages}{323} (\bibinfo{year}{2003}),

\bibitem[{\citenamefont{{Damour}}(1982)}]{1982-Damour-Proc}
\bibinfo{author}{\bibfnamefont{T.}~\bibnamefont{{Damour}}}, in
  \emph{\bibinfo{booktitle}{Marcel Grossmann Meeting: General Relativity}},
  edited by \bibinfo{editor}{\bibfnamefont{R.}~\bibnamefont{{Ruffini}}}
  (\bibinfo{year}{1982}).

\bibitem[{\citenamefont{{Price} and {Thorne}}(1986)}]{1986-Price.Thorne-PRD}
\bibinfo{author}{\bibfnamefont{R.~H.} \bibnamefont{{Price}}} \bibnamefont{and}
  \bibinfo{author}{\bibfnamefont{K.~S.} \bibnamefont{{Thorne}}},
  \bibinfo{journal}{Phys. Rev. D.} \textbf{\bibinfo{volume}{33}},
  \bibinfo{pages}{915} (\bibinfo{year}{1986}).

\bibitem[{\citenamefont{Thorne et~al.}(1986)\citenamefont{Thorne, Price, and
  Macdonald}}]{1986-Thorne.etal-Membrane}
\bibinfo{author}{\bibfnamefont{K.~S.} \bibnamefont{Thorne}},
  \bibinfo{author}{\bibfnamefont{R.~H.} \bibnamefont{Price}}, \bibnamefont{and}
  \bibinfo{author}{\bibfnamefont{D.~A.} \bibnamefont{Macdonald}},
  \emph{\bibinfo{title}{The Membrane Paradigm}}
  (\bibinfo{publisher}{Yale University Press}, \bibinfo{year}{1986}).

\bibitem{Parikh:1997ma}
  M.~Parikh and F.~Wilczek,
  Phys.\ Rev.\ D {\bf 58} (1998) 064011
  doi:10.1103/PhysRevD.58.064011
  [gr-qc/9712077].

\bibitem[{\citenamefont{{Policastro} et~al.}(2002)\citenamefont{{Policastro},
  {Son}, and {Starinets}}}]{2002-Policastro.etal-JHEP}
\bibinfo{author}{\bibfnamefont{G.}~\bibnamefont{{Policastro}}},
  \bibinfo{author}{\bibfnamefont{D.~T.} \bibnamefont{{Son}}}, \bibnamefont{and}
  \bibinfo{author}{\bibfnamefont{A.~O.} \bibnamefont{{Starinets}}},
  \bibinfo{journal}{Journal of High Energy Physics}
  \textbf{\bibinfo{volume}{9}}, \bibinfo{eid}{043} (\bibinfo{year}{2002}),
  \eprint{hep-th/0205052}.

\bibitem[{\citenamefont{{Bhattacharyya}
  et~al.}(2008)\citenamefont{{Bhattacharyya}, {Minwalla}, {Hubeny}, and
  {Rangamani}}}]{2008-Bhattacharyya.etal-JHEP}
\bibinfo{author}{\bibfnamefont{S.}~\bibnamefont{{Bhattacharyya}}},
  \bibinfo{author}{\bibfnamefont{S.}~\bibnamefont{{Minwalla}}},
  \bibinfo{author}{\bibfnamefont{V.~E.} \bibnamefont{{Hubeny}}},
  \bibnamefont{and}
  \bibinfo{author}{\bibfnamefont{M.}~\bibnamefont{{Rangamani}}},
  \bibinfo{journal}{Journal of High Energy Physics}
  \textbf{\bibinfo{volume}{2}}, \bibinfo{eid}{045} (\bibinfo{year}{2008}),
  \eprint{0712.2456}.

\bibitem[{\citenamefont{{Padmanabhan}}(2011)}]{2011-Padmanabhan-PRD}
\bibinfo{author}{\bibfnamefont{T.}~\bibnamefont{{Padmanabhan}}},
  \bibinfo{journal}{Phys. Rev. D.} \textbf{\bibinfo{volume}{83}},
  \bibinfo{eid}{044048} (\bibinfo{year}{2011}), \eprint{1012.0119}.

\bibitem[{\citenamefont{Kolekar and
  Padmanabhan}(2012)}]{2012-Kolekar.Padmanabhan-PRD}
\bibinfo{author}{\bibfnamefont{S.}~\bibnamefont{Kolekar}} \bibnamefont{and}
  \bibinfo{author}{\bibfnamefont{T.}~\bibnamefont{Padmanabhan}},
  \bibinfo{journal}{Phys. Rev. D} \textbf{\bibinfo{volume}{D85}},
  \bibinfo{pages}{024004} (\bibinfo{year}{2012}), \eprint{1109.5353}.

\bibitem[{\citenamefont{Bredberg et~al.}(2012)\citenamefont{Bredberg, Keeler,
  Lysov, and Strominger}}]{2012-Bredberg.etal-JHEP}
\bibinfo{author}{\bibfnamefont{I.}~\bibnamefont{Bredberg}},
  \bibinfo{author}{\bibfnamefont{C.}~\bibnamefont{Keeler}},
  \bibinfo{author}{\bibfnamefont{V.}~\bibnamefont{Lysov}}, \bibnamefont{and}
  \bibinfo{author}{\bibfnamefont{A.}~\bibnamefont{Strominger}},
  \bibinfo{journal}{J. High Energ. Phys.} \textbf{\bibinfo{volume}{2012}}
  (\bibinfo{year}{2012}). 

\bibitem[{\citenamefont{Chapline et~al.}(2003)\citenamefont{Chapline, Hohlfeld,
  Laughlin, and Santiago}}]{2003-Chapline.etal-IJMPA}
\bibinfo{author}{\bibfnamefont{G.}~\bibnamefont{Chapline}},
  \bibinfo{author}{\bibfnamefont{E.}~\bibnamefont{Hohlfeld}},
  \bibinfo{author}{\bibfnamefont{R.~B.} \bibnamefont{Laughlin}},
  \bibnamefont{and} \bibinfo{author}{\bibfnamefont{D.~I.}
  \bibnamefont{Santiago}}, \bibinfo{journal}{Int. J. Mod. Phys.}
  \textbf{\bibinfo{volume}{A18}}, \bibinfo{pages}{3587} (\bibinfo{year}{2003}),
  \eprint{gr-qc/0012094}.

\bibitem[{\citenamefont{Dvali and Gomez}(2012)}]{2012-Dvali.Gomez-PLB}
\bibinfo{author}{\bibfnamefont{G.}~\bibnamefont{Dvali}} \bibnamefont{and}
  \bibinfo{author}{\bibfnamefont{C.}~\bibnamefont{Gomez}},
  \bibinfo{journal}{Physics Letters B} \textbf{\bibinfo{volume}{716}},
  \bibinfo{pages}{240 } (\bibinfo{year}{2012}). 

\bibitem[{\citenamefont{{Dvali} and {Gomez}}(2014)}]{2014-Dvali.Gomez-EPJC}
\bibinfo{author}{\bibfnamefont{G.}~\bibnamefont{{Dvali}}} \bibnamefont{and}
  \bibinfo{author}{\bibfnamefont{C.}~\bibnamefont{{Gomez}}},
  \bibinfo{journal}{European Physical Journal C} \textbf{\bibinfo{volume}{74}},
  \bibinfo{eid}{2752} (\bibinfo{year}{2014}), \eprint{1207.4059}.

\bibitem[{\citenamefont{{Skakala} and
  {Shankaranarayanan}}(2016)}]{2016-Skakala.Shankaranarayanan-IJMPD}
\bibinfo{author}{\bibfnamefont{J.}~\bibnamefont{{Skakala}}} \bibnamefont{and}
  \bibinfo{author}{\bibfnamefont{S.}~\bibnamefont{{Shankaranarayanan}}},
  \bibinfo{journal}{International Journal of Modern Physics D}
  \textbf{\bibinfo{volume}{25}}, \bibinfo{eid}{1650047} (\bibinfo{year}{2016}).

\bibitem[{\citenamefont{Lopez et~al.}(2015)\citenamefont{Lopez, Bhattacharya,
  and Shankaranarayanan}}]{2015-Lopez.etal-Arx}
\bibinfo{author}{\bibfnamefont{J.~L.} \bibnamefont{Lopez}},
  \bibinfo{author}{\bibfnamefont{S.}~\bibnamefont{Bhattacharya}},
  \bibnamefont{and}
  \bibinfo{author}{\bibfnamefont{S.}~\bibnamefont{Shankaranarayanan}}
 \bibinfo{journal}{Phys. Rev.} \textbf{\bibinfo{volume}{D94}},
  \bibinfo{pages}{024029} (\bibinfo{year}{2016}),
 \eprint{1512.03680}.

\bibitem[{\citenamefont{Bhattacharya and
  Shankaranarayanan}(2014)}]{2014-Bhattacharya.Shankaranarayanan-Arx}
\bibinfo{author}{\bibfnamefont{S.}~\bibnamefont{Bhattacharya}}
  \bibnamefont{and}
  \bibinfo{author}{\bibfnamefont{S.}~\bibnamefont{Shankaranarayanan}}
  (\bibinfo{year}{2014}), \eprint{1411.7830}.

\bibitem[{\citenamefont{{Kubo}}(1966)}]{1966-Kubo-RPP}
\bibinfo{author}{\bibfnamefont{R.}~\bibnamefont{{Kubo}}},
  \bibinfo{journal}{Reports on Progress in Physics}
  \textbf{\bibinfo{volume}{29}}, \bibinfo{pages}{255} (\bibinfo{year}{1966}).

\bibitem[{\citenamefont{Bhattacharya and
  Shankaranarayanan}(2016)}]{2016-Bhattacharya.Shankaranarayanan-PRD}
\bibinfo{author}{\bibfnamefont{S.}~\bibnamefont{Bhattacharya}}
  \bibnamefont{and}
  \bibinfo{author}{\bibfnamefont{S.}~\bibnamefont{Shankaranarayanan}},
  \bibinfo{journal}{Phys. Rev.} \textbf{\bibinfo{volume}{D93}},
  \bibinfo{pages}{064030} (\bibinfo{year}{2016}), \eprint{1511.01377}.

\bibitem[{\citenamefont{Kulkarni et~al.}(1988)\citenamefont{Kulkarni,
  Chellathurai, and Dadhich}}]{1988-Kulkarni.etal-CQG}
\bibinfo{author}{\bibfnamefont{R.}~\bibnamefont{Kulkarni}},
  \bibinfo{author}{\bibfnamefont{V.}~\bibnamefont{Chellathurai}},
  \bibnamefont{and} \bibinfo{author}{\bibfnamefont{N.}~\bibnamefont{Dadhich}},
  \bibinfo{journal}{Class. Quant. Grav.} \textbf{\bibinfo{volume}{5}},
  \bibinfo{pages}{1443} (\bibinfo{year}{1988}),

\bibitem[{\citenamefont{Chellathurai and
  Dadhich}(1990)}]{1990-Chellathurai.Dadhich-CQG}
\bibinfo{author}{\bibfnamefont{V.}~\bibnamefont{Chellathurai}}
  \bibnamefont{and} \bibinfo{author}{\bibfnamefont{N.}~\bibnamefont{Dadhich}},
  \bibinfo{journal}{Class. Quant. Grav.} \textbf{\bibinfo{volume}{7}},
  \bibinfo{pages}{361} (\bibinfo{year}{1990}),

\bibitem[{\citenamefont{Dray et~al.}(1992)\citenamefont{Dray, Kulkarni, and
  Manogue}}]{1992-Dray.etal-GRG}
\bibinfo{author}{\bibfnamefont{T.}~\bibnamefont{Dray}},
  \bibinfo{author}{\bibfnamefont{R.}~\bibnamefont{Kulkarni}}, \bibnamefont{and}
  \bibinfo{author}{\bibfnamefont{C.~A.} \bibnamefont{Manogue}},
  \bibinfo{journal}{General Relativity and Gravitation}
  \textbf{\bibinfo{volume}{24}}, \bibinfo{pages}{1255} (\bibinfo{year}{1992}),

\bibitem[{\citenamefont{Delfino et~al.}(1996)\citenamefont{Delfino, Mussardo,
  and Simonetti}}]{1996-Delfino-NPB}
\bibinfo{author}{\bibfnamefont{G.}~\bibnamefont{Delfino}},
  \bibinfo{author}{\bibfnamefont{G.}~\bibnamefont{Mussardo}}, \bibnamefont{and}
  \bibinfo{author}{\bibfnamefont{P.}~\bibnamefont{Simonetti}},
  \bibinfo{journal}{Nuclear Physics B} \textbf{\bibinfo{volume}{473}},
  \bibinfo{pages}{469 } (\bibinfo{year}{1996}). 

\bibitem[{\citenamefont{Witten}(1984)}]{1984-Witten-CMP}
\bibinfo{author}{\bibfnamefont{E.}~\bibnamefont{Witten}},
  \bibinfo{journal}{Communications in Mathematical Physics}
  \textbf{\bibinfo{volume}{92}}, \bibinfo{pages}{455} (\bibinfo{year}{1984}). 

\bibitem[{\citenamefont{Shelton et~al.}(1996)\citenamefont{Shelton, Nersesyan,
  and Tsvelik}}]{1996-Shelton-PRB}
\bibinfo{author}{\bibfnamefont{D.~G.} \bibnamefont{Shelton}},
  \bibinfo{author}{\bibfnamefont{A.~A.} \bibnamefont{Nersesyan}},
  \bibnamefont{and} \bibinfo{author}{\bibfnamefont{A.~M.}
  \bibnamefont{Tsvelik}}, \bibinfo{journal}{Phys. Rev. B}
  \textbf{\bibinfo{volume}{53}}, \bibinfo{pages}{8521} (\bibinfo{year}{1996}),

\bibitem{Padmanabhan:2013nxa} 
  T.~Padmanabhan,
  Gen.\ Rel.\ Grav.\  {\bf 46}, 1673 (2014.

\bibitem[{\citenamefont{Beig}(1978)}]{1978-Beig-PLA}
\bibinfo{author}{\bibfnamefont{R.}~\bibnamefont{Beig}},
  \bibinfo{journal}{Physics Letters A} \textbf{\bibinfo{volume}{69}},
  \bibinfo{pages}{153 } (\bibinfo{year}{1978}). 

\bibitem[{\citenamefont{Szabados}(2004)}]{2004-Szabados-LRR}
\bibinfo{author}{\bibfnamefont{L.~B.} \bibnamefont{Szabados}},
  \bibinfo{journal}{Living Reviews in Relativity} \textbf{\bibinfo{volume}{7}}
  (\bibinfo{year}{2004}),

\bibitem[{\citenamefont{Wald}(1984)}]{1984-Wald-GeneralRelativity}
\bibinfo{author}{\bibfnamefont{R.~M.} \bibnamefont{Wald}},
  \emph{\bibinfo{title}{General Relativity}} (\bibinfo{publisher}{University of
  Chicago press}, \bibinfo{year}{1984}).

\bibitem[{\citenamefont{Tangherlini}(1963)}]{1963-Tangherlini-NC}
\bibinfo{author}{\bibfnamefont{F.~R.} \bibnamefont{Tangherlini}},
  \bibinfo{journal}{Il Nuovo Cimento (1955-1965)}
  \textbf{\bibinfo{volume}{27}}, \bibinfo{pages}{636} (\bibinfo{year}{1963}).

\bibitem[{\citenamefont{{Gourgoulhon}}(2005)}]{2005-Gourgoulhon-PRD}
\bibinfo{author}{\bibfnamefont{E.}~\bibnamefont{{Gourgoulhon}}},
  \bibinfo{journal}{Phys. Rev. D} \textbf{\bibinfo{volume}{72}},
  \bibinfo{eid}{104007} (\bibinfo{year}{2005}), \eprint{gr-qc/0508003}.

\bibitem[{\citenamefont{Gourgoulhon and
  Jaramillo}(2006)}]{2006-Gourgoulhon-PRD}
\bibinfo{author}{\bibfnamefont{E.}~\bibnamefont{Gourgoulhon}} \bibnamefont{and}
  \bibinfo{author}{\bibfnamefont{J.~L.} \bibnamefont{Jaramillo}},
  \bibinfo{journal}{Phys. Rev.} \textbf{\bibinfo{volume}{D74}},
  \bibinfo{pages}{087502} (\bibinfo{year}{2006}), \eprint{gr-qc/0607050}.

\bibitem[{\citenamefont{Kleinert and Shabanov}(1997)}]{1997-Kleinert-PLA}
\bibinfo{author}{\bibfnamefont{H.}~\bibnamefont{Kleinert}} \bibnamefont{and}
  \bibinfo{author}{\bibfnamefont{S.~V.} \bibnamefont{Shabanov}},
  \bibinfo{journal}{Phys. Lett.} \textbf{\bibinfo{volume}{A232}},
  \bibinfo{pages}{327} (\bibinfo{year}{1997}), \eprint{quant-ph/9702006}.

\bibitem[{\citenamefont{Landau and
  Lifshitz}(1980)}]{1980-Landau.Lifshitz-CourseofTheoretical}
\bibinfo{author}{\bibfnamefont{L.}~\bibnamefont{Landau}} \bibnamefont{and}
  \bibinfo{author}{\bibfnamefont{E.~M.} \bibnamefont{Lifshitz}},
  \emph{\bibinfo{title}{Statistical Physics, Pt. 1}} (\bibinfo{publisher}{Pergamon Press, London},
  \bibinfo{year}{1980}).

\bibitem[{\citenamefont{Bekenstein and
  Mukhanov}(1995)}]{1995-Bekenstein.Mukhanov-PLB}
\bibinfo{author}{\bibfnamefont{J.~D.} \bibnamefont{Bekenstein}}
  \bibnamefont{and} \bibinfo{author}{\bibfnamefont{V.~F.}
  \bibnamefont{Mukhanov}}, \bibinfo{journal}{Phys. Lett.}
  \textbf{\bibinfo{volume}{B360}}, \bibinfo{pages}{7} (\bibinfo{year}{1995}),
  \eprint{gr-qc/9505012}.

\bibitem[{\citenamefont{Hod}(1998)}]{1998-Hod-PRL}
\bibinfo{author}{\bibfnamefont{S.}~\bibnamefont{Hod}}, \bibinfo{journal}{Phys.
  Rev. Lett.} \textbf{\bibinfo{volume}{81}}, \bibinfo{pages}{4293}
  (\bibinfo{year}{1998}), \eprint{gr-qc/9812002}.

\bibitem[{\citenamefont{Dreyer}(2003)}]{2003-Dreyer-PRL}
\bibinfo{author}{\bibfnamefont{O.}~\bibnamefont{Dreyer}},
  \bibinfo{journal}{Phys. Rev. Lett.} \textbf{\bibinfo{volume}{90}},
  \bibinfo{pages}{081301} (\bibinfo{year}{2003}), \eprint{gr-qc/0211076}.

\bibitem[{\citenamefont{Skakala and
  Shankaranarayanan}(2014)}]{2014-Skakala.Shankaranarayanan-PRDa}
\bibinfo{author}{\bibfnamefont{J.}~\bibnamefont{Skakala}} \bibnamefont{and}
  \bibinfo{author}{\bibfnamefont{S.}~\bibnamefont{Shankaranarayanan}},
  \bibinfo{journal}{Phys.Rev. D} \textbf{\bibinfo{volume}{89}},
  \bibinfo{pages}{044019} (\bibinfo{year}{2014}), \eprint{1311.4255}.

\bibitem{Kothawala:2008in} 
  D.~Kothawala, T.~Padmanabhan and S.~Sarkar,
  Phys.\ Rev.\ D {\bf 78}, 104018 (2008)
  [arXiv:0807.1481 [gr-qc]].

\bibitem[{\citenamefont{Wheeler}(1992)}]{1992-Wheeler-Proc}
\bibinfo{author}{\bibfnamefont{J.~A.} \bibnamefont{Wheeler}}, in
  \emph{\bibinfo{booktitle}{Proceedings of Sakharov memorial lectures}}, edited by
  \bibinfo{editor}{\bibfnamefont{L.}~\bibnamefont{Keldysh}} \bibnamefont{and}
  \bibinfo{editor}{\bibfnamefont{V.}~\bibnamefont{Feinberg}}
  (\bibinfo{publisher}{Nova}, \bibinfo{year}{1992}).

\bibitem[{\citenamefont{Kawai et~al.}(1987)\citenamefont{Kawai, Lewellen, and
  Tye}}]{1987-Kawai.Lewellen.Tye-NPB}
\bibinfo{author}{\bibfnamefont{H.}~\bibnamefont{Kawai}},
  \bibinfo{author}{\bibfnamefont{D.~C.} \bibnamefont{Lewellen}},
  \bibnamefont{and} \bibinfo{author}{\bibfnamefont{S.-H.~H.}
  \bibnamefont{Tye}}, \bibinfo{journal}{Nuclear Physics B}
  \textbf{\bibinfo{volume}{288}}, \bibinfo{pages}{1 } (\bibinfo{year}{1987}). 

\bibitem[{\citenamefont{Bern}(2002)}]{2002-Bern-LRR}
\bibinfo{author}{\bibfnamefont{Z.}~\bibnamefont{Bern}},
  \bibinfo{journal}{Living Rev. Rel.} \textbf{\bibinfo{volume}{5}},
  \bibinfo{pages}{5} (\bibinfo{year}{2002}), \eprint{gr-qc/0206071}.

\bibitem[{\citenamefont{Ananth and Theisen}(2007)}]{2007-Ananth.Theisen-PLB}
\bibinfo{author}{\bibfnamefont{S.}~\bibnamefont{Ananth}} \bibnamefont{and}
  \bibinfo{author}{\bibfnamefont{S.}~\bibnamefont{Theisen}},
  \bibinfo{journal}{Phys. Lett.} \textbf{\bibinfo{volume}{B652}},
  \bibinfo{pages}{128} (\bibinfo{year}{2007}).



\bibitem{Fischler:2015cma}
  W.~Fischler and S.~Kundu,
  Phys.\ Rev.\ D {\bf 92} (2015) no.4,  046008
  doi:10.1103/PhysRevD.92.046008
  [arXiv:1501.01316 [hep-th]].
	
\bibitem{Fischler:2015kro}
  W.~Fischler and S.~Kundu,
  JHEP {\bf 1604} (2016) 112
  doi:10.1007/JHEP04(2016)112
  [arXiv:1512.01238 [hep-th]].
	

\bibitem[{\citenamefont{Sykes et~al.}(2012)\citenamefont{Sykes, Haar, and
  Pitaevskii}}]{2012-Lifshitz-Perspectives}
\bibinfo{author}{\bibfnamefont{J.}~\bibnamefont{Sykes}},
  \bibinfo{author}{\bibfnamefont{D.}~\bibnamefont{Haar}}, \bibnamefont{and}
  \bibinfo{author}{\bibfnamefont{L.}~\bibnamefont{Pitaevskii}},
  \emph{\bibinfo{title}{Perspectives in Theoretical Physics: The Collected
  Papers of E$\backslash$M$\backslash$Lifshitz}} (\bibinfo{publisher}{Elsevier
  Science}, \bibinfo{year}{2012}). 

\bibitem[{\citenamefont{P.~W.~Atkins}(2006)}]{1999-Atkins.dePaula-physicalchemistry}
\bibinfo{author}{\bibfnamefont{J.~P.~d.} \bibnamefont{P.~W.~Atkins}},
  \emph{\bibinfo{title}{Physical Chemistry, $8^{th}$ edition}}
  (\bibinfo{publisher}{W. H. Freeman}, \bibinfo{year}{2006}).

\bibitem[{\citenamefont{Pauling}(1935)}]{1935-Pauling-JACS}
\bibinfo{author}{\bibfnamefont{L.}~\bibnamefont{Pauling}},
  \bibinfo{journal}{Journal of the American Chemical Society}
  \textbf{\bibinfo{volume}{57}}, \bibinfo{pages}{2680} (\bibinfo{year}{1935}).

\bibitem[{\citenamefont{Chow and Wu}(1987)}]{1987-Chow.Wu-PRB}
\bibinfo{author}{\bibfnamefont{Y.}~\bibnamefont{Chow}} \bibnamefont{and}
  \bibinfo{author}{\bibfnamefont{F.~Y.} \bibnamefont{Wu}},
  \bibinfo{journal}{Phys. Rev. B} \textbf{\bibinfo{volume}{36}},
  \bibinfo{pages}{285} (\bibinfo{year}{1987}),

\end{thebibliography}
\end{document}